%% file: main-with-names.tex
\documentclass[sigconf,nonacm]{acmart}
\renewcommand\footnotetextcopyrightpermission[1]{} 
\AtBeginDocument{%
  }

\setcopyright{none} 
\acmConference[ ]{ }{ }{ }

\usepackage{circuitikz}
\usepackage{pgfplots}
\usepackage{pgfplotstable}
\usepackage{subcaption}
\usepackage{multirow}
\usepackage{array}
\usepackage{tikz}
\usepackage{bookmark}

\newcolumntype{C}[1]{>{\centering\arraybackslash}p{#1}}

\begin{document}

\title{HaShiFlex: A High-Throughput Hardened Shifter DNN Accelerator with Fine-Tuning Flexibility}

\author{Jonathan Herbst}
\affiliation{
  \institution{Brown University}
  \country{USA}
}

\email{jonathan\_herbst@brown.edu}

\author{Michael Pellauer}
\affiliation{
  \institution{NVIDIA}
  \country{USA}
}
\email{mpellauer@nvidia.com}

\author{Sherief Reda}
\affiliation{
  \institution{Brown University}
  \country{USA}
}

\email{sherief\_reda@brown.edu}







\begin{abstract}

  We introduce a high-throughput neural network accelerator that embeds most network layers directly in hardware, minimizing data transfer and memory usage while preserving a degree of flexibility  via a small neural processing unit for the final classification layer. By leveraging power-of-two (Po2) quantization for weights, we replace multiplications with simple rewiring, effectively reducing each convolution to a series of additions. This streamlined approach offers high-throughput, energy-efficient processing, making it highly suitable for applications where model parameters remain stable, such as continuous sensing tasks at the edge or large-scale data center deployments. Furthermore, by including a strategically chosen reprogrammable final layer, our design achieves high throughput without sacrificing fine-tuning capabilities. 
  
  We implement this accelerator in a 7nm ASIC flow using MobileNetV2 as a baseline and report throughput, area, accuracy, and sensitivity to quantization and pruning---demonstrating both the advantages and potential trade-offs of the proposed architecture. We find that for MobileNetV2, we can improve inference throughput by $20\times$ over fully programmable GPUs, processing 1.21 million images per second through a full forward pass while retaining fine-tuning flexibility. If absolutely no post-deployment fine tuning is required, this advantage increases to $67\times$ at 4 million images per second.

\end{abstract}



\maketitle

\section{Introduction}





Hardware accelerators have become a cornerstone of modern deep learning, largely because of the computational intensity associated with training and inference in neural networks. As datasets continue to grow and model architectures become more complex, general-purpose processors such as CPUs often struggle to deliver the high throughput and energy efficiency necessary for real-time or large-scale applications. By contrast, specialized hardware designs can dramatically reduce the time and energy required to execute neural network operations.

Specialized architectures built for neural network processing come in several broad categories. First, there are GPU-based solutions, which combine Turing-complete streaming multiprocessors with integrated tensor cores to accelerate the matrix operations commonly found in neural network workloads. Second, there are FPGA-based accelerators that gain efficiency through reconfigurable logic, allowing researchers to tailor the hardware pipeline to specific neural architectures. Third, specialized ASICs like Google’s TPU and various neural processing units are designed from the ground up to address specific patterns of computations in neural networks. Each approach offers distinct trade-offs in terms of flexibility, power efficiency, and raw computational capabilities. Each of these solutions store the neural network weight parameters in dynamically writable RAM memories that are loaded with the weight values from the most recent version of the model at runtime. Multi-layer neural networks are time-multiplexed through the accelerator hardware using the best-known scheduling policy, updating the weight memories with the relevant parameters for each layer.

In contrast, hardwiring the accelerator by fixing weights directly into the hardware can substantially simplify the underlying architecture. A fixed-weight design can eliminate all weight related memory transfer overheads by embedding learned parameters in logic. Furthermore, dedicating hardware resources to only those operations necessary for the fixed model can streamline data pathways via design-time constant propagation, further increasing performance. To accomplish this, different layers of the network are no longer time-multiplexed into the same datapaths and RAMs, but instead are spatially ``unrolled'' into different physical areas of the chip. Thus, while per-layer area efficiency increases, the overall area required to provision an entire DNN can be quite large. On the other hand, this spatial distribution of resources also results in a high-throughput pipelined execution across layers, and so the tradeoff can be favorable.

Hardwired accelerators offer compelling advantages for scenarios requiring high-performance, energy-efficient inference with relatively static or infrequently updated models. They are particularly well-suited for applications where the target model is stable for extended periods and speed or power efficiency is paramount—such as embedded systems performing continuous sensing or computer vision tasks, edge devices with stringent power budgets, or large-scale data centers running fixed workloads at massive scale. This combination of efficiency, reliability, and performance can enable real-time responsiveness and broader deployment possibilities, especially in domains where frequent retraining is not necessary or where hardware upgrade cycles naturally accommodate incremental updates to the accelerator design. Of course, the natural downside of this is that post-deployment hardware adaptability decreases, which can be a large problem in a fast-moving field like Deep Learning.

In this paper, we propose a high-throughput accelerator architecture that simultaneously addresses the downsides of overall area efficiency and hardware adaptability when using hardwired weights. 
To retain a degree of post design-time flexibility, we include a small neural processing unit (NPU) dedicated to handling the final classification layer, thus enabling on-the-fly adjustment of model outputs (i.e., fine-tuning). 
Furthermore, we leverage power-of-two quantization for hardwired weights, allowing us to dispense with multipliers or shifters entirely, as each shift operation becomes a simple hardware re-wiring. This dramatically increases per-layer area efficiency, thus making our design \emph{the first to make full spatial unrolling of a realistic neural network feasible in a limited area budget}. It is our hope that our combination of hardening and flexibility will inspire future research in this rich design space.
We summarize the specific contributions of this paper as follows:

\begin{itemize}
 \vspace{0.07in}    
\item We present a novel hardwired CNN accelerator design that eliminates the need for storing and transferring weights. By quantizing weights to powers of two, all multiplication operations are replaced with rewiring, thereby reducing convolutions to a sequence of additions. We also show how batch normalization and non-linear activations can be seamlessly integrated into this framework.
\vspace{0.07in}    
\item To preserve flexibility, we incorporate a small neural processing unit (NPU) that handles the final classification layer in a traditional manner. This approach offers adaptability in the output stage, mitigating some of the rigidity introduced by hardwiring the earlier layers.
 \vspace{0.07in}   
 \vspace{0.07in}    
\item We analyze the area implications of our proposed architecture and identify widely-used CNN models that fit within reticle constraints, demonstrating how they can be effectively deployed to leverage the benefits of hardwired processing.
  \vspace{0.07in}       
\item Using MobileNetV2 as a baseline and implementing our design in a 7 nm ASIC flow, we provide a comprehensive set of experimental results evaluating throughput, area, accuracy, and flexibility. We also conduct a sensitivity analysis to assess how bit quantization and pruning affect these performance metrics.
\end{itemize}

 The remainder of this paper is organized as follows. In the Background section, we review key concepts and existing accelerator architectures relevant to our proposed design. Next, in Section 3 we detail our hardwired architecture, including the quantization strategy and the integration of a small neural processing unit for added flexibility. In Section 4 we discuss training. In Section 5, we present comprehensive performance benchmarks, including throughput, area, accuracy, and sensitivity analyses. We then discuss our work within the broader research landscape in the Related Work section, discussing key differences and similarities to prior accelerator designs. Finally, Section 7, summarizes our main findings and suggests possible directions for future research.

\section{Background}

\subsection{Convolutional Neural Networks}
One of the most important types of Deep Neural Networks (DNNs) is the Convolutional Neural Network (CNN), formalized by Yann LeCun as a way to extract spatial information from images \cite{LeCun:CNN}. CNNs ``slide'' a filter over the previous layer's features, using several filters to separate information into planar ``channels.''

Throughout this paper, we use the following convention, defined in Eyeriss \cite{Eyeriss:ISCA}: $P$ and $Q$ represent the convolutional output's height and width, $M$ represents output channels, $R$ and $S$ represent the kernel's height and width, and $H$, $W$, and $C$ represent the convolutional input's height, width, and number of channels respectively. In addition, we use $O$ or $O^{P \times Q \times M}$ to represent the convolution output, $W$ or $W^{M \times R \times S \times C}$ to represent the filter matrix, and $I$ or $I^{H \times W \times C}$ to represent the convolution input. Thus we can write each output element in tensor algebra summation notation:

$$O_{pqm} = \sum_{r=0}^R \sum_{s=0}^S\sum_{c=0}^C W_{m,r,s,c} \cdot I_{p+r,q+s,c}$$ 

In practice, many accelerators map this operation onto a 2-dimensional matrix multiplication, known as the Toeplitz representation. We flatten the $M$ filter weight matrices $W^{R \times S \times C}$ into the 2D matrix $W^{M \times RSC}$, and unroll the input image $PQ$ times to get $X^{RSC \times PQ}$. Thus, another equivalent way to write the equation would be $$O^{PQ \times M} = W^{M \times RSC} X^{RSC \times PQ}$$ after reshaping the results to a 3-dimensional format.

In order to calculate an output element, we need RSC filter weights and inputs, which must be loaded from memory. Because hardware devices only have limited on-chip storage, these values must be continuously streamed from main memory, which provides a memory bottleneck that reduces computational intensity. 

\subsection{Hardware-Friendly Quantization \& Pruning}

In order to reduce these memory costs (as well as increase area- and energy-efficiency) we can compress the model size through two techniques known as \emph{quantization} and \emph{pruning}. Quantization reduces the number of bits used to represent each weight, which sometimes also applies to the activations; this can also involve changing the representation type, for instance from floating-point to fixed-point. This applies across all weights in the model, although some layers can have different quantization than other layers, i.e., mixed-precision \cite{Chen_2021_ICCV, wang2019haq, wu2018mixed}.

\begin{figure}[t]
\resizebox{\columnwidth}{0.1\textheight}{\input{graphs/weight-encoding}}
\caption {\small 2:4 Weight Compression. We slice the matrix along its row axis in groups of four elements, then extract the two nonzero elements into a smaller matrix. Our metadata matrix contains two-bit indices of the nonzero values in their original groups.}
\end{figure}
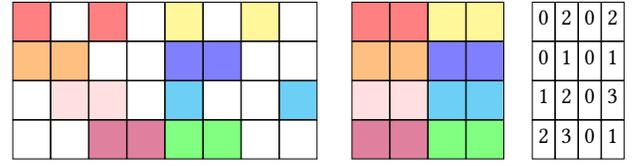

\begin{table*}[t]
\centering
\caption{Qualitative comparison to existing DNN accelerators. See Figure \ref{fig:size_sweep} for detailed quantification of area unrolling costs.}
\begin{scriptsize}
\begin{tabular}{|l|l|c|c|C{1.3in}|C{0.8in}|C{0.5in}|c|c|}
\hline
\multirow{2}{*}{\textbf{Design Category}} & \multirow{2}{*}{\textbf{Design}} & \multicolumn{2}{c|}{\textbf{Weight Value Strategy}} & \multirow{2}{*}{\textbf{Datapath Strategy}} & \multirow{2}{*}{\textbf{Multi-Layer Strategy}} & \multirow{2}{*}{\textbf{Throughput$^\dagger$}} & \multicolumn{2}{c|}{\textbf{Qualitative Area Cost}} \\
\cline{3-4}\cline{8-9}
 & & \textbf{Design-Time} & \textbf{Run-Time} & & & & \textbf{Per Layer} & \textbf{Unrolled} \\
\hline
\hline
\multirow{4}{*}{PE Array} & Eyeriss \cite{Eyeriss:ISCA} & - & Dynamic & Multipliers (Row Stationary) & Time-Multiplexing & Medium & Full Die & \textbf{Infeasible} \\
\cline{2-9}
 & DianNao \cite{DianNao} & - & Dynamic & Multipliers (Weight Stationary) & Time Multiplexing & Medium & Full Die & \textbf{Infeasible} \\
\cline{2-9}
 & TPU \cite{TPU:ISCA} & - & Dynamic & Multipliers (Systolic Array) & Time Multiplexing & Medium & Full Die & \textbf{Infeasible} \\
\hline
\hline
\multirow{1}{*}{Hardened PE Array} & \emph{Nonexistent} & Fixed & - & Constant-Propagated Multipliers & Space Unrolling & High & Medium & \textbf{Infeasible} \\
\hline
\hline
\multirow{3}{*}{Shifter-Based} & ShiftCNN \cite{ShiftCNN} & - & Dynamic Po2 & Shift-and-Add (ShiftALU)& Time Multiplexing & Medium & Medium & \textbf{Infeasible} \\
\cline{2-9}
& Universal Shift \cite{UniversalShift} & - & Dynamic Po2 & Shift-and-Add (conv2d-shift) & Time Multiplexing & Medium & Medium & \textbf{Infeasible} \\
\cline{2-9}
& Jumping Shift \cite{JumpingShift} & - & Dynamic Po2 & Shift-and-Add (2-bit barrel) & Time Multiplexing & Medium & Medium & \textbf{Infeasible} \\
\cline{2-9}
\hline
\hline
\multirow{2}{*}{\emph{\textbf{This Work}}} & \emph{HaShiFix} & Fixed Po2 & - & Rewire (Constant-Propagated Shift) & Space Unrolling & Highest & Very Low & Large \\
\cline{2-9}
 & \emph{HaShiFlex} & Fixed Po2 & Fine Tuning & Rewire (Constant-Propagated Shift) and Multipliers & Space Unrolling and Time Multiplexing & High & Low & Full Die \\
\hline
\end{tabular}
\label{tab:dnn_accelerators}

\begin{flushright}
$^\dagger$ Refers to throughput of completing a forward pass through all DNN layers.
\end{flushright}
\end{scriptsize}
\end{table*}

The other major compression technique is pruning, which eliminates some proportion of the model's weights entirely. We can perform either \emph{structured} or \emph{unstructured} pruning, where the former involves clamping weights in some organized pattern and the latter allows us to clamp any of the weights without regard to organization. Both pruning formats reduce the number of multiplications needed, since we can ignore the zero weights, but they have limited benefits in many accelerators. In particular, GPUs compress in a 2:4 weight compression strategy, which preserves two weights out of every four elements in a row group. This means transferring half the weights and inputs instead of the full amount, but also requires us to transfer a compression matrix with the indices of each nonzero element (Figure 1). This means that for matrices $W^{PQ\times RSC}$ and $X^{RSC\times M}$, instead of transferring PQRSC + RSCM weights we transfer $PQ\frac{RSC}{2}$ + $\frac{RSC}{2}M$ weights and $PQ\frac{RSC}{2}$ 2-bit indices in our metadata matrix.

In terms of compute cycles, 2:4 weight compression allows us to run a smaller matrix multiplication: instead of $W^{M\times RSC} X^{RSC \times PQ}$ we only compute $W^{M\times\frac{RSC}{2}}  X^{\frac{RSC}{2}\times PQ}$, where the inner dimension is only halfsize. We model this cycle count with SCALE-Sim \cite{samajdar2020systematic} in our results section to understand sparsity's effect on accelerator throughput. 

\subsection{Contrast to Existing DNN Accelerators}

DNN accelerators fall into three subtypes: fully programmable (i.e., GPUs), fixed function multiplier arrays (i.e,, TPUs), and Field-Programmable or Coarse-Grained Reconfigurable Arrays (also called NPUs). Each of them has a main characteristic in common, which is a loadable array of processing elements, or PEs. To execute a complete forward pass through a multi-layer Deep Neural Network, these PEs load the parameters of each layer in sequence (i.e., time multiplexing of the hardware datapaths), passing output features from one layer into the next via on-chip storage. 

Table \ref{tab:dnn_accelerators} shows a detailed breakdown of why these PE approaches cannot be used directly to achieve our goals. Namely, while they achieve maximum post-deployment ability to change DNN parameters, overall forward pass throughput suffers because of this time multiplexing.
Thus, when field updating of weights is not required, it is natural to investigate whether using design-time constant weight values and constant-propagation is feasible to increase throughput. This approach requires
spatially ``unrolling'' the hardware for each layer -- as each layer has different fixed weight values. Unfortunately, multipliers themselves are large and area inefficient. This style of constant propagation does not result in sufficient area savings to allow for fitting an entire realistic sized DNN on a reticle-limited die. 

One discovery in the field of energy-efficient computing is that it can often be sufficient to simply shift the activation; in other words, instead of multiplying X by 4, we simply left-shift it by 2. Our compute unit thus reduces from a multiplier-accumulator to a shifter-accumulator, where $X_{kj}$ is shifted by $\log_2(W_{ik})$. This works best when our weight values are already \textbf{power-of-two (Po2)}, and some research has been conducted on making neural networks accurate when quantized to Po2 weights \cite{zhou2017, DeepShift}, but the hardware works regardless of weight values. Because shifters are much smaller than multipliers, our computational density becomes much higher, allowing us to achieve a higher throughput for the same area cost. Most work done with shift accelerators is performed academically, and is run on FPGAs or CGRAs for cost-purposes; the general terminology for such an array is Neural Processing Unit or NPU, which refers to a systolic array of specific hardened computational units. We refer to these specific shifting array designs as ``Shift NPUs,'' since they are systolic arrays with shift-and-accumulate compute units. For a detailed discussion of these approaches see Section \ref{sec:related}.

Our key insight is that hardening design-time fixed Po2 weight values into rewiring-and-accumulate modules results in a significant jump in per-layer area efficiency. As we describe in detail in the next section, this both makes the spatial unrolling of a realistic size neural network feasible, and also allows enough area left over for supporting strategic flexibility. 
    
    
    
\section{Proposed Hardwired Accelerator Architecture}

\begin{figure*}
\resizebox{0.9\textwidth}{!}{\includegraphics[]{graphs/Full-Chip-Overview-5.pdf}}
\caption{\small An overview of the HaShiFlex ASIC design. Image inputs are loaded into on-chip memory, run through the hardcoded convolution layers, and the feature extractor outputs are stored in an intermediate buffer. These activations are run through a flexible NPU array that multiplies them against stored on-chip weights to compute the final classification probabilities, which are streamed off-chip.}
\end{figure*}



In our architecture, given in Figure 2, we load the initial inputs into on-chip memory using high-speed interconnect, run the image through our hardened convolution, and save the results in a small buffer that reflects the outputs of our feature extractor. By using power-of-two weight compression, which reduces the multiplication to rewiring when the convolution weights are fixed, we are able to fit our entire neural network architecture on-chip. The last layer – a fully connected layer for classification – is run in an on-chip NPU to maintain classification flexibility, and the outputs are fed back to our main processor. Our chip's input is a series of HxW color images streamed onto the chip, and the output is a series of k-length vectors that correspond to softmax probabilities which are streamed back out from the chip. Our hardened convolution corresponds to the feature extractor of neural network architectures, while the on-chip NPU is the classifier layer that accesses the feature-extractor outputs ($k_{fe}$) stored in registers.

We begin by explaining the benefits of this hardcoded approach, explain each of the subcomponents' translation to hardware in greater detail, and conclude the section with a description of potential tradeoffs of the fixed-function design.

Using a hardened accelerator comes with several benefits, enumerated below.

\subsubsection{No weight transfers from off-chip} Because our chip ``stores'' the convolution weights in the operation of its convolution engine, we do not need to continually fetch and store our filter weights from DRAM. Off-chip memory access comes with a hefty latency and energy pricetag; thus, by avoiding this bottleneck we can achieve both higher throughput and higher energy efficiency, a combination which is often a tradeoff in other settings.    

\subsubsection{No inter-layer I/O transfer} In addition to not needing to transfer weights from DRAM, we also do not suffer from the difficulty that other DNN accelerators face, namely, transferring the outputs of each layer either off-chip or into a streaming buffer. Because all layers of our design happen simultaneously on-chip, every layer's inputs and outputs do not need to be intermediately stored but propagate through the design. This only requires us to store the last outputs in an on-chip buffer, which is orders of magnitude smaller ($k_{fe}$ instead of a double-buffer with both layer inputs and outputs i.e. $PQM + HWC$; for example, in MobileNetV2 this is 1280 elements compared to 802,816 elements to store Layer 2 Conv) and thus requires both less energy and area to store. 

\subsubsection{Fast cycle counts and throughput for entire model}

The main benefit of our hardcoded design is its cycle count. Because we constant-propagate all of our convolution operations in parallel, and because each convolution operation involves spatial accumulation instead of accumulating over PEs, our entire feature extractor's latency reduces to several cycles. This allows extremely high throughput, on the order of millions of images per second (Section 5), making it a good design choice for users with high throughput demand.

\subsubsection{Increased area efficiency} Because our convolutional layers are purely combinational, we greatly reduce the amount of area and energy required to calculate and store our elements (Section 5). In particular, our shifting-to-rewiring algorithm removes the need for multiplication or even shifters, but translates directly to rewiring, which uses negligible area and energy. This translates to immediate cost savings, allowing users to purchase a greater number of chips at a functionally equivalent running-price.

\subsubsection{Sparsity has immediate benefits} There has been much research on increasing sparsity in convolutional neural networks without decreasing accuracy \cite{xu2020convolutional}. However, this has limited benefits in the actual hardware of most accelerators: sparse weight matrices can be compressed, but their reliance on compression schemes requires structured sparsity to run efficiently, which limits our search space for sparsity benefits. In addition, because traditional accelerators must still compute matrices of reduced size, the decrease in our compute cycles is actually sublinear (Section 5). In contrast, pruning an operation from a hardened, constant-propagated accelerator means removing its rewire-and-accumulate from the chip; this does not require compression schemes because the compression format is hardwired onto the chip, allowing for benefits at any pruning percentage, and the isomorphism between sparsity and hardware units dictates a linear reduction in area and energy. 

\subsection{Hardwired Convolution}

Our neural network compression depends on quantizing our weights to power-of-two format. The traditional algorithm for a convolution uses floating-point 32-bit weights:

$$O_{pqm} = \sum_{r=0}^R \sum_{s=0}^S\sum_{c=0}^C W_{m,r,s,c} \cdot I_{p+r,q+s,c}$$

where a traditional accelerator involves a processor array of full 32-bit multipliers to compute the element-wise multiplication. How the accelerator divides up its multipliers for element computation depends on chosen stationarity, but involves time-multiplexing the same multipliers for multiple computations.

We rely on DeepShift's \cite{DeepShift} quantized activations and weights, where our weights are power-of-two format. When we represent $W_{m,r,s,c}$ as $2^p$ for some integer p within bitshift range, the multiplication simplifies to a bit-wise shift:

$$W_{m,r,s,c} x = 2^\text{p}x = \left\{\begin{array}{lr}
        x << \text{p} & \text{if p} > 0\\
        x >> \text{p} & \text{if p} < 0\\
        x  & \text{if p} = 0
        \end{array} \right. $$

Each power-of-two quantization entails some individual rounding errors (Figure 3) but the network on the whole can achieve state-of-the-art precision for image classification \cite{DeepShift}. 

Previous works have utilized the hardware-friendly shifter operation to reduce area-size, where the chip setup is equivalent to a ``Shift NPU'' with processing arrays of variable shifters in place of multipliers \cite{Multiplierless, UniversalShift, WithoutMultiplication}. Our crucial difference from this approach involves the realization that, at inference-time, the neural network's filter weights are fixed; therefore, if we can propagate all of our multipliers on-chip instead of time-multiplexing them, then each weight multiplication reduces to a fixed shifter, where the shifter is tied to the specific filter weight. However, this is equivalent to a simple rewiring, which takes no area at all. Moreover, there is no accuracy tradeoff to this calculation, because the result of rewiring is equivalent to the result of a fixed-function shifter, which is in turn equivalent to the result of a binary shifter set to a specific value at runtime. 

\begin{figure*}
    \centering
    \begin{subfigure}[b]{0.35\textwidth}
        \centering
        \resizebox{\textwidth}{!}{\input{graphs/fig1.1}}
        \caption{Stage 1; FP32 on TPU}
    \end{subfigure}
    \hspace{3em}
    \begin{subfigure}[b]{0.35\textwidth}
        \centering
        \resizebox{\textwidth}{!}{\input{graphs/fig1.2}}
        \caption{Stage 2; Fixed-point Po2 on TPU}
    \end{subfigure}
    
    \vspace{1em} 
    
    \begin{subfigure}[b]{0.35\textwidth}
        \centering
        \resizebox{\textwidth}{!}{\input{graphs/fig1.3}}
        \caption{Stage 3; ShiftCNN}
    \end{subfigure}
    \hspace{3em}
    \begin{subfigure}[b]{0.35\textwidth}
        \centering
        \resizebox{\textwidth}{!}{\input{graphs/fig1.4}}
        \caption{Stage 4; Rewire-and-Accumulate}
    \end{subfigure}
    
    \caption{The four stages of multiplication compression. We begin with a full n-bit multiplier, which involves $O(n^2)$ full adders and $O(n)$ half-adders. A traditional accelerator time-multiplexes this multiplier, with the weights being refreshed for each compute. Stage two involves converting the weights to the nearest power-of-two, which incurs some accuracy reduction. Stage three replaces our multiplier with a bit shifter, which uses $O(n)$ multiplexers. Shift NPUs time-multiplex this shifter in the same way that traditional accelerators time-multiplex their multipliers. Finally, stage 4 is our novel approach: by constant-propagating all shifters on-chip and hardening their weight values, we convert each shifter into a simple rewiring, which uses no logic.}
\end{figure*}
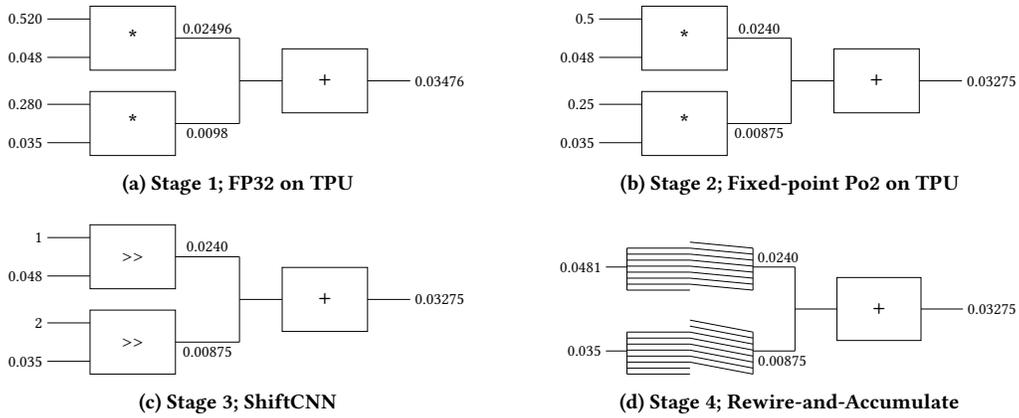

Because we rewire instead of using multipliers, our area overhead for each element multiplication is nonexistent, so we can constant propagate all of our values to happen at the same time. The area cost for our convolution is solely due to the reduction of these elements into one output: this requires $RSC-1$ adders, which may further simplify (i.e., addition of two right-shifted elements would require only a 7-bit adder instead of an 8-bit one). 

To prevent overflow, we need an expanding adder tree such that our first layer is eight bits, our second layer is nine bits, and so on. Each layer halves the number of partial sums that need to be reduced, which halves the number of adders of that size required. We have RSC computations in each partial sum reduction to a single output $O_{pqm}$; our number of layers is the number of times we have to halve RSC before reducing to a single output, which is $log_2(RSC)$. We have $\frac{RSC}{2}$ 8-bit additions in the first layer to add all RSC elements together; in the successive layer we need $\frac{RSC}{4}$ 9-bit additions, and this number halves with each layer. Thus, our adder tree's area is proportional to the following formula:

 $$\sum_{i=0}^{log_2(RSC)} (RSC \cdot 2^{-(1+i)}) \text{\ (8+i)-bit adders}.$$

If we reduce our activation quantization to n-bits, this can have a significant saving in area and energy by reducing the size of our adders; we analyze this tradeoff in Section 5.2. 

The large number of adders used in our constant-propagated design also influenced our choice of algorithms: another commonly used quantization scheme for neural networks is additive-power-of-two or APOT weights, where each weight is represented as an addition of several powers of two to smooth the values. Each power of two inside of the weight would be computed as a separate input rewiring in our design, which corresponds to an extra full-adder; thus, using APOT with two sub-elements for each weight would double the size of the design, while three sub-elements would triple it. Given the near-SOTA accuracy for Po2 quantization, we elected not to utilize APOT for this design.

In total we have $PQM$ adder-trees for each layer, where the formula for the adder-trees' area is given above. The total number of elements is large, but as Table 4 shows, most adder-trees are fewer than a hundred microns and several hundred microns at maximum; thus, we will be able to fit our convolution on-chip for a carefully chosen model and pruning. 

\subsection{Batch Normalization}
The second main computation in neural networks is batch normalization, which does a per-channel normalization of the inputs:
$$batchnorm(x_i) = \gamma \left ( \frac{x_i - \mu_b}{\sqrt{\sigma_b^2 + \epsilon}} \right ) + \beta$$
where $\gamma$ and $\beta$ are learned scale and shift-parameters and $\mu_b, \sigma_b$ are the computed batch mean and variance, respectively. To compute these means and variances would require per-element subtraction, multiplication, and division, which would incur heavy area and power costs when performed in either a constant-propagated or time-multiplexed manner. However, because our chip is only used for inference, the mean and variance are not actually computed but are used from stored running-buffers, which stay constant; thus, we can fold the weights into the convolution:

$$batchnorm_{inf}(x_i) = \gamma \left ( \frac{(Wx_i) - \mu_b}{\sqrt{\sigma_b^2 + \epsilon}} \right )+ \beta$$$$ = \left (\frac{\gamma W}{\sqrt{\sigma^2 + \epsilon}} \right ) x_i + \left (\beta - \frac{\gamma \mu}{\sqrt{\sigma^2 + \epsilon}} \right )$$

which is the same as a biased convolution operation. PikeLPN \cite{Neseem2024PikeLPNMO} quantizes this batch normalization by quantizing $s = Q\left ( \frac{\gamma}{\sqrt{\sigma^2+\epsilon}} \right )$ and $b = Q(\beta) - Q \left (\frac{\gamma \mu}{\sqrt{\sigma^2 + \epsilon}} \right )$ and writes the batch normalization as $batchnorm(x) = x*s - b$, with the ability to fold the quantized batch norm into the convolution at runtime. Our quantization augments this approach with the additional constraint that our weights need to be quantized to Po2; therefore, we quantize all variables $\gamma, \beta, \sigma_b, \mu_b$ to fixed-point and  additionally quantize $\gamma$ and $\sqrt{\sigma^2+\epsilon}$ to a Po2 representation. Since all weights in W are Po2, we can combine them together to get a single shifting convolution that corresponds to the combined variance-normalized convolution, and then add a fixed-point bias term to offset the mean. This requires one additional adder for each output term, or $PQM$ additional adders per layer.

\subsection{ReLU}
Finally, we focus on our activation function. Several activation functions are commonly used in neural networks, including sigmoid, softmax, leaky ReLU, and Swish; we focus on ReLU, as it is the most common function and the cheapest to implement in hardware. ReLU keeps the value of its input if it is positive, otherwise it clamps it to zero:

$$ReLU(x) = \left\{\begin{array}{lr}
       x & \text{if x} > 0\\
        0 & \text{if x} \leq 0\\
        \end{array} \right.$$
        
With a signed fixed-point input, we can perform this in hardware by inverting the most significant bit and anding it with the rest of the bits. Because this operation only requires two gates, it is very small in hardware, requiring only 2 cells and 0.1 $\mu \text{m}^2$ area in Asap 7nm; it must be performed once for each output, or $PQM$ times per layer.

\subsection{Flexible Classification from Final Layer}
While our feature extractor is compressed and hardened on-chip, the last layer(s) of neural networks is a classifier, which is a fully-connected layer performing matrix-vector multiplication to compute the class logits for our input. If the feature extractor has $k_{fe}$ outputs and can be written as a vector $X^{k_{fe}}$ and our network is a single matrix multiplication A that outputs k classes, this is an $mc$ matrix multiplication
$$O^{m} = A^{m \times c}X^c.$$

To maintain flexibility and transfer learning capability, it is crucial that this layer is performed on an on-chip NPU, where the weights are kept in a local buffer and can be streamed from the processor if the user wants to perform a different task.

Leaving this last layer on-chip incurs negligible cost. While we explore the specific cycle count for MobileNetV2 in Section 5.2, the fact that this is a relatively small matrix-vector multiplication means we can perform this operation without incurring heavy area or latency costs. For instance, in MobileNetV2 the classification head is a 1000-by-1280 weight matrix multiplying a 1280-by-1 vector, which makes up only $0.4\%$ of the MACs performed across all of the feature extractor layers. Our design involves our feature extractor activations being fed directly into the NPU, allowing for an output-stationary dataflow with a cycle count proportional to k. In particular, we note in Section 5 that this cycle count is often comparable to the number of cycles required to load our input image data onto the chip, and thus by using pipelining we entirely amortize this cost. Both the weights and the inputs to the NPU are stored in local buffers, where the input buffer is the intermediate buffer used to store our feature extractor activations and the weights buffer is a $k \times k_{fe}$ element buffer that can also be accessed from the main memory bus. This allows the user to stream new transfer learning weights onto the chip, which are stored locally and allow the chip to perform the new image classification task. After we perform this final matrix multiplication, we gather the final output classification probabilities and stream them off-chip.

\subsection{Tradeoffs of Hardwired Accelerators}

Limited academic exploration of hardened CNN accelerators can largely be attributed to fears about the drawbacks of hardening a chip design, especially given the slow development time of ASICs and quick development time of new models. However, many of the initial drawbacks to the approach are not limiting:

\subsubsection{The design is size-limited} Users who are interested in achieving accuracy at the cost of any other variable might be dissuaded by the limitation of fitting the network entirely on-chip. ResNet-50 achieves maximal classification accuracy among the CNN models, but even compressed using our Po2-rewiring scheme its sheer number of elements and layers greatly exceeds reticle limit. However, we argue that ResNet is not necessary to achieve good accuracy; Google's MobileNetV2 and V3 architectures both achieve within 5\% accuracy with hundreds of times fewer operations, and remain the standard among lightweight models while also being competitive among full-scale models. Thus, the size-limitation to run lightweight models should not pose a downside for most use cases.

\subsubsection{The design is architecture-specific} Another concern is that by creating hardware that only encodes a specific model, which is likely to become outdated by the time the chip is manufactured. However, we note that both MobileNetV2 and MobileNetV3 have been around since 2019 and remain along the size-accuracy pareto frontier. We believe that the long duration of these models eases these concerns and allows for the chip's production without obsolescence factors.

\subsubsection{Quantization and sparsity required for sufficient compression} Finally, users might be concerned about the utilization of both Po2 quantization and sparsity in our model design. In particular, they might be concerned that Po2 quantized models cannot achieve SOTA for their use-cases, while other critics might argue that since the benefits of sparsity are so persuasive, there will be a race to continually sparsify the model, preventing it from ever being hardened for production. To the first case, we state that Po2 eight-bit quantized models have minimal accuracy loss when retrained on the relevant dataset \cite{DeepShift}, and we demonstrate in Section 5 that this also applies to transfer learning situations. To the second case, we acknowledge that our sparsity results may be surpassed; however, we claim that neural networks can only become so sparse before incurring an unacceptable accuracy loss, and that future attempts at pruning will not lead to instability in the specific die mask used long-term.

\section{Training the Model}

\subsection{Neural Network Architecture Selection}

To utilize our chip properly, we first need to carefully choose our model; using too large a neural network would prevent us from being able to harden the feature extractor on the chip, while using a small or obsolete model would limit our accuracy and use-cases. We performed an analysis of different popular CNN models, such as the MobileNet, VGGNet, and ResNet families, and plotted their feature extractor size when hardened on ASIC against their top-1 accuracy from the pretrained model zoo (Figure 4). 

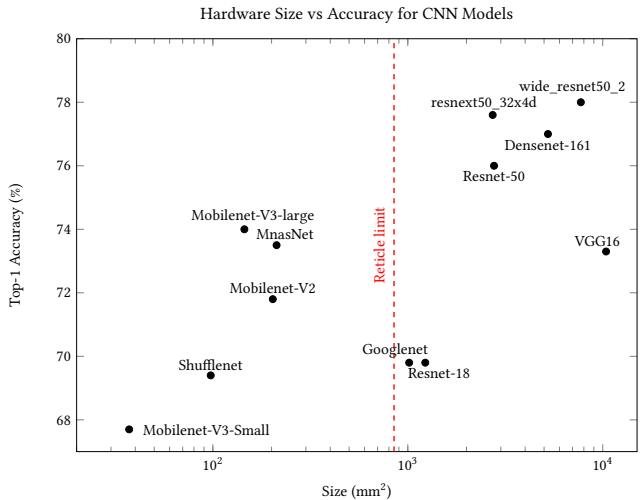
\begin{figure}
\centering
\resizebox{\columnwidth}{!}{\input{graphs/size-accuracy2}}
\caption{\small Neural network architecture size when quantized to Q3.5 and compressed using our hardcoding procedure, compared to top-1 accuracy in pretrained model zoo. The MobileNet family gives a good tradeoff of chip size without sacrificing accuracy.}
\label{fig:size_sweep}
\end{figure}

As we can see, there is a tradeoff between size and accuracy that prevents us from hardening ResNet-50, a very deep and dense model, as its hardened size is far above the reticle limit. However, the MobileNet families achieve high accuracies while being orders of magnitude smaller than ResNet, making them a good choice for our analysis. In particular, MobileNetV2 \cite{DBLP:journals/corr/abs-1801-04381} is conceptually easy to translate to hardware, as it consists only of convolutional layers, batch normalization, and ReLU activations while still achieving top-1 accuracy similar to VGG-19. While MobileNetV3-large achieves better accuracy than V2, its usage of Hardsigmoid and Hardswish activations requires lookup tables instead of the simple invert-and-and operation for ReLU, increasing both model complexity and area, such that MobileNetV2 remains the best candidate for this approach.

\subsection{Quantization and Secondary Sparsification}

For a given neural network, there are a variety of quantization approaches, including INQ \cite{zhou2017}, ABC-Net \cite{lin2017accuratebinaryconvolutionalneural}, and LogNN \cite{miyashita2016convolutionalneuralnetworksusing}, which take different approaches to compression and training regimes. Our hardware accelerator requires a Po2-weights compression with minimal encoding of each multiplication to a single binary shift to minimize adders, while quantizing our activations allows our adders to have lower bitwidths; thus, we used DeepShift \cite{DeepShift} as it provides all of these functionalities.

Existing research has shown that we can perform weight-Po2 quantization on Resnets and VGGNets with minimal Imagenet top-1 accuracy loss, and on MobileNetV2 with minimal CIFAR-10 top-1 accuracy loss \cite{DeepShift}. We begin by quantizing pretrained MobileNetV2 to verify that the model's Imagenet top-1 accuracy loss is minimal. We also quantize our inputs to fixed-point implementation to fit on-chip; we assess accuracy for different input bitwidths. Because our weight shifting corresponds to input rewiring, there is no hardware savings from reducing our weight bitwidth to 5 bits as performed in DeepShift. Therefore, instead of reducing our weight bitwidth, we keep this value equal to the value of our input bitwidth; for example, an input quantization of 3 integer bits and 5 fraction bits, represented Q3.5, would have a weight bitwidth of 8. Our default quantization was Q3.5 as performed in DeepShift \cite{DeepShift}, and all learning parameters were kept the same between quantizations. Our fully classified layer was kept at FP32, as this will be run at full-precision on an on-chip NPU.

An additional, novel operation to DeepShift's experimental paradigm involved quantizing our batch normalization layers. As shown in Section 3.2, our accelerator involves folding the batch normalization into the convolution operation, which requires the batchnorm layer's running mean and shift variables to be fixed-point quantized and the running mean and scale variables to be Po2 fixed-point quantized. We run our analysis with these quantizations, where the Po2 bitwidths correspond to the weight bitwidths in the convolutional layers and the fixed-point quantizations correspond to the input bitwidths into the convolution.

After identifying the bitwidth-models within our accuracy threshold, we use sparsity as a secondary compression technique, and assess the amount of pruning that each model can achieve without severely reducing the accuracy. In particular, while we will be folding our batch normalization layers into the convolution when hardening our inference engine, it is sufficient to prune our convolutional layers to achieve this sparsity savings. Because our convolution weights become $\left ( \frac{\gamma W}{\sqrt{\sigma^2 + \epsilon}} \right )$ after folding, the batch-normalization factor $s = \frac{\gamma}{\sqrt{\sigma^2 + \epsilon}}$ is a scalar multiplier onto our filter weights, and does not change which weights are smallest and should be pruned. Thus, we can prune just our convolutional layers and fold the batch normalization at runtime without consequences.

Because MobileNetV2's depthwise separable convolutions incur minimal area costs (Table 2), we do not sparsify  these layers nor the first layer of our model. We begin pruning from the previous, quantized model version and train at an initial learning rate of 0.001 decreasing over time, with monotonically increasing sparsity.



\section{Evaluation}

We first analyze the statistics of our accelerator, breaking down the area, latency, and power of the chip's feature extractor, classifier layer, and on-chip buffers. 

\textbf{Methodology:} We use Cadence Genus Synthesis Version 19.10-p002\_1 with the Asap7nm PDK, mapping to gate-level logic and optimizing. Our last-layer NPU is based on SCALE-Sim's \cite{samajdar2020systematic} verilog code. We use MobileNetV2 for our model, which takes in images of $224\times224\times3$ and has a feature extractor output size of 1280, which it classifies using a $1000 \times 1280$ matrix multiplication. We quantize MobileNetV2 to Q3.5 with Po2 weights, and train the model to 60\% sparsity with minimal accuracy loss (Section 5.2). 
We run SCALE-Sim in GEMM format for different NPU sizes and show the speedup factor for 2:4 weight encoding. Finally, we assess the flexibility of our model by employing transfer learning on other image datasets and comparing the results to transfer learning with the original MobileNetV2.

\subsection{Main Result: Area, Throughput, and Latency}

\begin{normalsize}
\begin{table}[h!]
  \centering
  \caption{Sub-component Area For Pruned MobileNetV2}
  \label{table:area}
  \small
  \begin{tabular}{l|l}
    Layer name & Size ($mm^2$) \\
    \hline
    \hline   
    Feature extractor & 219 \\
    \hline
    1024x1 on-chip NPU & 0.24 \\
    \hline
    On-chip buffers & 0.42 \\
    \hline
    Total (1 accelerator) & \textbf{220} \\
    \hline 
    Total (4 accelerators) & 880 \\
    \hline
  \end{tabular}
 
\end{table}
\end{normalsize}

Detailed area breakdown is shown in Table \ref{table:area}. Using $60 \%$ sparsity, our area for a Q3.5 feature extractor is 219 $\text{mm}^2$. On-chip buffers and our last layer NPU take up less than 0.5 $\text{mm}^2$. Running SCALE-Sim \cite{samajdar2020systematic} with a general matrix multiplication of 1000x1x1280 MNK output-stationary format, our minimum achievable cycle count is 2278 cycles on a 1000x1 NPU array. Using SCALE-Sim's provided systolic array code, our area costs for this unit is 0.24 $\text{mm}^2$. Thus, each individual accelerator is \textbf{220 $\text{mm}^2$}.
We can duplicate our accelerator four times before reaching reticle limit, for a total chip size of 880 $\text{mm}^2$.

In contrast, HaShiFix does not have this last layer NPU, and thus does not require either the multiplier cost or the 1280-element buffer used to separate the feature extractor from the classifier unit. However, as these are relatively small area costs compared to the size of our accelerator, our HaShiFix area is not significantly different from HaShiFlex.

 

\begin{table}[h!]
  \centering
  \caption{Comparison against SOTA Accelerators}
  \small
  \label{table:sota}
  \begin{tabular}{l|l|l|l}
    Design & Throughput (im/s) & Latency ($\mu s$) & Area ($\text{mm}^2$) \\
    \hline
    \hline
    \textbf{HaShiFlex} & 1.21 * $10^6$& 3.3 & 880 \\
    \hline
    \textbf{HaShiFix} & 4.0 * $10^6$& 0.25 & 550 \\
    \hline
    \hline
    H100 GPU & $\approx $60,000 & & 814 \\
    \hline
    Google TPU v4 & 100 & $\approx$ 2,600 & 600 \\
    \hline
    GraphCore M2000 & $\approx$ 10,000 & 520 & 4 $\times$ 823 \\
    \hline
  \end{tabular}
 
\end{table}

Table \ref{table:sota} shows an overall breakdown of throughput and latency for HaShiFlex and HashiFix derived from our methodology. Overall, we find our flexible HaShiFlex approach able to finish full forward passes through all layers on millions of images per second, an improvement of roughly $20.2\times$ in throughput over fully programmable GPU solutions. In situations where no post-deployment adaptability is needed, this advantage increases to $67\times$ for the fully-fixed weight HaShiFix solution.

\textbf{Comparative Analysis: } Numbers are hard to find for MobileNetV2 throughput and inference for datacenter TPUs and GPUs, but all metrics underperform by several orders of magnitude. Google's Edge TPU achieves 2.6 ms latency for unpruned eight-bit fixed-point MobileNetV2 \cite{coral_benchmarks}, which is \textbf{800x} slower than our design. Running on a server TPU, we found inference throughput of 100 FPS for an INT8 MobileNetV2 with a batch-size of 32, which is \textbf{12000x} below our design.  An NVIDIA H200 GPU can run ResNet-50v1.5 at INT8 at 65,045 images/sec \cite{nvidia_ai_inference}; while we cannot directly compare statistics for MobileNetV2, this indicates orders of magnitude less volume. Similarly, the GraphCore IPU-M2000 can achieve 0.52 ms latency and 9,404 images/sec throughput on Q16.16 ResNet-50v1.5 with configurations tuned specifically to optimize these parameters; even accounting for latency reduction for fixed-point 8-bit and MobileNet, both our throughput and latency perform 2 orders of magnitude better. Jiang et al. \cite{Mobilenetv2FPGA} created an FPGA accelerator specifically for Mobilenet-V2, which achieved 1910 FPS; our work achieves a \textbf{600x} better framerate. On MLPerf, a Qualcomm Hexagon (TM) NPU achieved 2,036.56 images/sec on MobileNetV4 \cite{mlcommons_inference_mobile}, which is again \textbf{600x} less volume. The smallest reported image classification latency for a MobileNet on MLPerf \cite{mlcommons_inference_tiny} was Qualcomm's Snapdragon 8 Gen 3 Mobile HDK, which achieved 190 $\mu s$ on MobileNetV1 0.25x, a much smaller model; even so, their latency is still \textbf{57x} larger than ours. Given the existing limitations of reticle area and interconnect speeds, our design is limited by memory bandwidth for a chip that only loads the image itself and outputs the final output; thus, it achieves the maximum achievable throughput and minimum latency for CNN inference.

\subsection{Efficiency Improvement Details}
Because these bottom-line numbers can be difficult to understand in isolation, we now engage in a thorough explanation of the source of the efficiency improvements.
 
\begin{itemize}
    \item Our sparsity linearly impacts the size of our chip. For a chip of size 549 $\text{mm}^2$, a sparsity factor of $s$ evaluates to a chip of size $a_{individ} = 549(1-s)$.
    \item Our parallelization factor depends on chip size; given a reticle limit of 850 $\text{mm}^2$, the amount of possible parallelization is $k = \left \lfloor{\frac{850}{a_{individ}}} \right \rfloor$.
    \item NVIDIA's H100 GPU is $814 \text{mm}^2$ and has an interconnect speed of 900 GB/s. Interconnect scales linearly with area size, giving us interconnect speeds of $900 * \frac{\text{area}}{814} = 900 * \frac{k*a_{individ}}{814}$, but each accelerator can only access $\frac1k$ of the memory bandwidth, giving each accelerator a memory bandwidth of $\frac{900}{814}a_{individ}$ GB/s.
    \item Our chip operates at 1 GHz, limited by our NPU array; our feature extractor operates at a much slower speed, but does not need to be clocked at the same speed. Our per-chip interconnect corresponds to $\frac{900}{814}a_{individ}$ bytes of data bus.
    \item For each run, an accelerator needs to load one image and store one 1000-length vector of probabilities off-chip. Since our accelerator operates at Q3.5, we only need to load 8-bit image vectors, but our final classification probabilities are FP32, requiring 32 bits. The amount of data through memory bandwidth is $(224^2*3 + 1000*8)$ bytes per accelerator.
    \item Our NPU array requires 3300 cycles, regardless of sparsity or area sizes. Because our data input, feature extractor, and NPU are pipelined, our maximum latency is determined by the larger of 3300 cycles and the number of cycles to load our data. Below $65\%$ sparsity, we are limited by NPU cycle count; above this sparsity, we are limited by data loading. 
    \item Our latency directly corresponds to our cycle count; for 3300 cycles, we achieve \textbf{3.3 $\mu$ s latency}.
    \item Our throughput is equal to our parallelization factor multiplied by our individual accelerator throughput, which is $k\frac{10^9}{l}$ (Figure 5c). For $65\%$ sparsity, we achieve \textbf{1.2 million images/sec}. Our maximum throughput is at $69\%$ sparsity, where we achieve 1.24 million images/sec. Beyond this point, the effects of limited memory bandwidth outweigh the benefits of parallelization, and both throughput and latency degrade.
\end{itemize}

In contrast, HaShiFix is not limited by the 3300 cycle count, but only by the time to load the data. Thus, we calculate our cycles directly from parallelization and data load time, included in Figure 5c below. At $0\%$ sparsity, we have the maximum chip size allowing for the most memory bandwidth; $a_{individ} = 549 \text{mm}^2$, so our interconnect speed is 607 GB/s. With a 607-byte bus devoted to a single accelerator loading $(224^2*3 + 1000*8)$ bytes of data, this only takes 250 cycles, allowing for a latency of \textbf{$0.25 \mu s$}, 13 times lower than our HaShiFlex latency. Because we can't parallelize, our throughput becomes $\frac{10^{9}}{250}$ = \textbf{4 million} images/sec, almost four times higher than HaShiFlex's design.    

\subsection{Impact of Model Compression}

Given that our model is originally unpruned and only quantized to eight-bit, we assess the extent to which we can compress our model along both quantization and sparsity axes, as well as the size of the model in each case. 

\begin{normalsize}
\begin{table}[h!]
  \centering
  \caption{Hardened Convolution sizes ($\mu m^2$) for different input bitwidths in Asap 7nm.}
  \label{table:formatting}
  \small
  \begin{tabular}{l|l|l|l|l}
    Convolution & 5-bit & 6-bit & 7-bit & 8-bit \\
    \hline
        \hline

    3x3x3 & 27.3 (54.6\%) & 35.9 (71.8\%)	& 43.5 (87\%) & 50.0 \\
    \hline
    3x3 (pw) & 1.0 (100.0\%) & 1.0 (100.0\%) &	1.0 (100.0\%) & 1.0 \\
    \hline
    1x1x16 & 16.4	(55.8\%) & 21.0 (71.4\%) & 25.0 (85\%) &	29.4 \\
    \hline
    1x1x32 & 33.3 (54.6\%)	& 43.7 (71.6\%) & 50.0 (82\%) & 61.0 \\
    \hline
    1x1x64 & 72.6	(57.6\%) & 88.2 (70.0\%) & 106.4 (84.4\%) & 126.0\\
    \hline
    1x1x320 & 362.9 (57.4\%) & 450.7 (71.2\%) & 543.2 (85.9\%) & 632.6 \\
    \hline
    MAC Unit & 16.8 (53.9\%) & 21.3 (68.3\%) & 25.9 (83.0\%) & 31.2 \\
    \hline
  \end{tabular}
 
\end{table}
\end{normalsize}

\begin{normalsize}
\begin{table}[h!]
  \centering
  \caption{MobileNetV2 Weight bit (WB) Po2 Quantization and Input bit (IB) Quantization Versus Accuracy}
  \label{table:formatting}
  \small
  \begin{tabular}{l|l|l}
    Input Bitwidth & Top-1 Accuracy (\%) & Top-5 Accuracy (\%) \\
    \hline
    \hline
    Original & 72 & 90 \\
    \hline
    WB 8, IB Q3.5 & 68 & 88 \\
    \hline
    WB 7, IB Q3.4 & 66 & 87 \\
    \hline
    WB 7, IB Q2.5 & 55 & 80 \\
    \hline
    WB 6, IB Q3.3 & 56 & 80 \\
    \hline
    WB 6, IB Q2.4 & 22 & 45 \\
    \hline
    WB 5, IB Q3.2 & 0.23 & 1.1 \\
    \hline
  \end{tabular}
\end{table}
\end{normalsize}

Running our convolution modules in Genus with the Asap7 PDK, we get the area costs for an entire hardened convolution unit (Table 4). The majority of our area costs are in pointwise convolutions, which are also our best candidates for compression. Reducing our activations from 8-bit to 7-bit can provide area savings of $15\%$, while compressing to five bits is roughly half the area cost. However, when running MobileNetV2 with different input bitwidths, our accuracy degradation quickly becomes infeasible. Notably, our results degrade much more quickly than in DeepShift \cite{DeepShift} because we quantize our activations as well as our weights, as activation quantization is crucial for reducing our adder bitwidths, while weight quantization (due to being simple rewiring) does not affect our area metrics. We find a sharp accuracy decline between Q3.5 and Q3.4 quantizations and all other quantizations; in addition, only Q3.5 is within a 5\% accuracy threshold. Thus we use Q3.5 as our baseline quantization scheme.

\begin{figure*}
\centering
\begin{subfigure}[b]{0.3\textwidth}
    \centering
    \resizebox{\textwidth}{!}{
    \input{graphs/sparsity-vs-accuracy}}
    \caption{Accuracy vs Sparsity}
\end{subfigure}
\hfill
\begin{subfigure}[b]{0.3\textwidth}
    \centering
    \resizebox{\textwidth}{!}{
    \input{graphs/area_vs_sparsity}}
    \caption{Area vs Sparsity}
\end{subfigure}
\hfill
\begin{subfigure}[b]{0.3\textwidth}
    \centering
    \resizebox{\textwidth}{!}{
    \input{graphs/throughput_vs_sparsity}}
    \caption{Throughput vs Sparsity}
\end{subfigure}
\hfill
\caption{Analysis of the effects of pruning on accuracy, area, and throughput. Accuracy analysis is taken from pruning our Q3.5 MobileNetV2 model with incremental sparsity of 20\% and 30 epochs of retraining. Area analysis is taken from translating the resulting sparse convolution modules into Genus with Asap7nm. Throughput analysis comes from Section 5.1 and depends on achievable interconnect speed and parallelization capacity.}
\end{figure*}
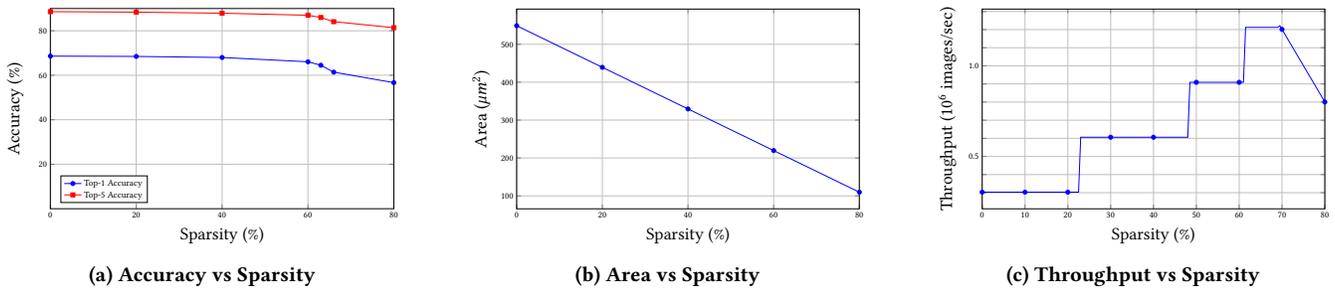


Pruning is particularly beneficial for our chip, since sparsity level linearly decreases our area (Figure 5b). It is therefore critical to identify the maximal amount of pruning possible before our accuracy decreases. Running incremental pruning of $20\%$, with 30 layers of post-pruning retraining and with each pruning occurring on top of the previous fully-retrained model, we find that we can prune to $60\%$ before significantly degrading our accuracy. Running a secondary, more fine-grained pruning of $3\%$ incremental pruning with 10 epochs retraining from $60\%$ to $69\%$ sparsity threshold, we find that our accuracy degrades quickly after reaching $60\%$ sparsity, so we utilize this value for our hardwired design. This level of sparsity places us at 219 $\text{mm}^2$, which allows us a four-times parallelization factor and corresponding throughput increase. 

In contrast, GPUs can only compress in a 2:4 weight compression strategy, which preserves two weights out of every four elements, or a 50\% compression. Systolic arrays in general are often not built to take advantage of the small convolution sizes in MobileNetV2; TPU v4, for instance, is built around 128x128 systolic arrays and does not dynamically resize them, meaning that the arrays are already underutilized for smaller pointwise convolutions such as $1\times 1 \times 24$. However, when we run the layers in SCALE-Sim \cite{samajdar2020systematic} on a 128-by-128 weight-stationary config file, mirroring the TPU systolic arrays, we do find that the array can take advantage of 2:4 weight encoding. When we compare the cycle counts for each layer of MobileNetV2 against the same layer when the inner dimension is halfsize, $W^{M\times \frac{RSC}{2}} X^{\frac{RSC}{2} \times PQ}$ compared to $W^{M\times RSC} X^{RSC \times PQ}$, we find an average cycle count reduction to $83\%$ of the original cycle count across the different layers, resulting in $60\%$ of the total cycles. Depending on the specific layer choice, therefore, we can achieve linear savings on layers like 1x1x160 and 1x1x24, but fail to achieve cycle savings on badly tiled layers such as 1x1x144 or 1x1x192. In general, we note that while neural networks can be carefully constructed to be responsive to 2:4 weight encoding, the overall savings are often sublinear to the amount pruned.

In general, our hardware design is far more friendly to pruning than other accelerators: because pruning a weight from a layer directly corresponds to removing its adder from the reduction tree, our hardware demonstrates benefits when we prune to any amount. As we linearly reduce our area with sparsity, we can increasingly parallelize our accelerator, with no drawbacks to latency since we continue to be compute-bound by our individual last-layer NPUs. 

\subsection{Post-Deployment Flexibility and Accuracy}

In order to make our model competitive against other inference accelerators, it is crucial that we remain able to adapt our feature extractor to different related image classification tasks. Without this capability, our accelerator's use-cases would be limited to baseline ImageNet classification; however, by retraining our last layer we can demonstrate successful transfer learning onto new datasets.

In particular, we compare our quantized, sparsified version of MobileNetV2 to the original MobileNetV2 for accuracy in transfer learning the CIFAR-10 and CIFAR-100 datasets. We note, crucially, that HaShiFix has a hardened classifier layer, making it impossible to perform transfer learning when the task is different from ImageNet; HaShiFlex's last layer flexibility allows it to be useful across a much broader range of tasks.

Our analysis is performed on the original FP32 MobileNetV2, the Q3.5 HaShiFlex model without pruning, and the Q3.5 HaShiFlex model pruned to $60\%$ sparsity. The last layer is kept at FP32 precision and fully programmable, while all other layers are frozen. We chose Flowers102 \cite{Nilsback08}, OxfordPets \cite{parkhi12a}, and Food101 \cite{bossard14} as three representative datasets of transfer learning regimes, with high input resolution that benefit strongly from CNN's feature extraction capabilities. In addition, we included CIFAR10 and CIFAR100 \cite{krizhevsky2009learning} as Transfer Learning regimes, since their low resolution can demonstrate the limits of nonresizable neural networks. We retrained our models for 20 epochs with an initial learning rate of 0.01 and a step size of 5 epochs.


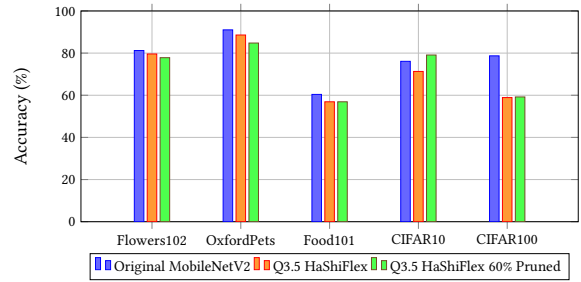
\begin{figure}
\resizebox{0.9\columnwidth}{!}{\input{graphs/transfer-learning}}
\caption{\small Transfer Learning Accuracy Across Image Classification Datasets. All models are MobileNetV2 with fixed input dimension and frozen except for the last layer. Original is pretrained from model zoo, Quantized is our 8-bit Po2 MobileNetV2 without sparsity, and Sparse is pruned to $60\%$ sparsity.}
\end{figure}


We see that our hardened model maintains sufficient flexibility in its last layer to learn new image classification tasks. All three variants suffer significantly for CIFAR10 and CIFAR100, since these are only 32-by-32 pixel images and our accelerator can only work with 224-by-224 inputs; thus, rescaling the image to this size causes most learned spatial features to be rendered useless. However, we believe that there are few cases when images will need to be processed at such low resolutions, and our datasets Oxford Pets and Stanford Dogs occur in variable image sizes, demonstrating that our network is still effective at learning from downsampled images. 

Thus, we can see that our accelerator achieves throughput speedups of 400-600 relative to other hardware, with latency reduction on the order of 1000x. This makes it an interesting alternative for high-throughput and low-latency inference environments.

\section{Related Work}
\label{sec:related}

\subsection{Shifter-Based Hardware Accelerators}

In 2017 Godovskiy and Rigazio introduced ShiftCNN \cite{ShiftCNN}, which proposed increasing hardware computational efficiency by using power-of-two weights in order to remove the multipliers and replace them with shifters. Furthermore, ShiftCNN employed aggressive post-training quantization that enabled pre-computation of all possible terms in the output tensor based on the weights. Although this work was never formally peer reviewed, we include it because it had a large influence on the remaining works in this section.

Hsieh et al. propose a multiplier-less CNN accelerator \cite{Multiplierless} based around programmable RAM-based lookup tables whose sizes are limited by using a custom 4-bit quantization format called Signed Digit 4 (SD4). Beyond post-training quantization, the authors developed several techniques to aid in the challenges of in-training quantization, including compensating for gradient deltas that were notably smaller than the SD4 format's range would encode.

Song et al. propose a ``universal shift'' CNN accelerator \cite{UniversalShift} based on 8-bit activations and 4-bit power-of-two weights using lookup tables. Training is accomplished using shifted representation in-training, with post-training quantization.

Xu et al. propose a non-uniform codebook quantization using Huffman encoding \cite{WithoutMultiplication} that can be implemented via lookup tables. They accelerate this pattern using a hierarchical systolic array of 14x14 PEs, where each PE consists of a 5-element 1D systolic shifter.
\vspace{-15pt}

\subsection{Multiplication-Free Neural Networks}

Inspired by hardware accelerator efficiency gains, a category of work looks to leverage those gains by exposing the lack of multiplication operation directly to the neural network.

ShiftNet \cite{Shift} described CNN convolutions by enumerating all the possible ``shift matrices'' that describe the shift directions of a layer. This allowed all multiplications to be replaced with pointwise memory transfers. Unfortunately, this made memory bandwidth a bottleneck. The Sparse Shift Layer \cite{SparseShiftLayer} attempts to reduce this by eliminating ineffectual shifts, particularly when paired with quantization.

DeepShift \cite{DeepShift} is a software-only neural network that uses shift-based techniques directly in the architecture of the layers. Additional contributions included reducing the number of shifts in a convolutional layer from ShiftCNN's 2-3 to 1, and extending the technique to linear layers. Additionally, DeepShift demonstrated that training power-of-two weights from scratch was possible, in addition to post-training quantization.

ShiftAddNet \cite{ShiftAddNet} uses separate shift and add layers, and demonstrated that the anchor-weight phenomenon can exist in shifting layers as well as traditional multiplied weights. AddressNet \cite{AddressNet} specializes these into three primitives called ``channel shift,'' ``address shift,'' and ``shortcut shift'' that have efficient GPU implementations.

In contrast, AdderNet \cite{AdderNet} demonstrated that shifts could be removed altogether by maximizing emphasis on addition operations, replacing the convolutional dot-product with calculation of the $\ell_1$ distance between filter and input.

PikeLPN \cite{Neseem2024PikeLPNMO} is a network based around the insight that non-quantized elementwise operations such as activation functions become the bottleneck. It proposes a technique called QuantNorm to quantize batch norm layers that we leverage and expand upon.

\subsection{Shifter-Based Quantization}

Finally, we discuss a survey of methods aimed at improving accuracy for power-of-two-based quantization accuracy. We include this work as an overview of techniques that demonstrate the feasibility of our approach, and could be combined with our insights as future work.

DenseShift \cite{DenseShift} combines power-of-two quantization with sparse pruning by proposing zero-free shifting. The resulting weights are densely encoded without zeros (which is technically not a power-of-two). Accuracy is recovered using a sign-scale decomposition technique during training.

JumpingShift \cite{JumpingShift} proposes an alternative scheme called Jumping Logic Quantization wherein a new coefficient is added to power-of-2 exponents, effectively ``jumping'' over regions in the quantization range, and also demonstrate removing zeros to increase shifter utilization without loss of accuracy.

Global sign-based network quantization (GSNQ) \cite{LogQuant} assigns different ranges based on the sign bit of the weights during training. The authors demonstrate competitively high accuracy by retraining CNNs both inter-layer and intra-layer levels, effectively allowing later layers to compensate for pruned damage to earlier layers.

S$^3$ \cite{S3} observes that shift-based networks are highly sensitive to weight initialization during training, and can further suffer from vanishing gradients or sign-bit freezing due to short bit widths. They propose a sign-sparse-shift decomposition representation that improves learning rate and decreases initialization sensitivity.

\section{Discussion}

Our novel approach to full constant propagation of large neural networks allows us to reduce our area costs in significant ways. Crucially, because all weights exist on-chip, we can utilize power-of-two representation to convert each necessary multiplication into a fixed bit shifter and thus ultimately to rewiring. Thus, our area costs only come from the adder trees necessary for partial sum accumulation of outputs, and can fit in 220 $\text{mm}^2$ for a lightweight, sparse network such as sparsified MobileNetV2. Our approach allows for complete input multicasting and reuse, and eliminates the need to transfer any data on- or off-chip besides the input data, thus achieving the theoretical maximum inference speed on hardware.

Because of the state-of-the-art accuracy results of MobileNets even when these models are sparse, we are able to reduce our inference engine to fit comfortably on-chip and to allow for easy parallelization for increased throughput. Unlike GPUs and TPUs, which approach the reticle limit and still don't achieve relatively high model throughput, our design opens the possibility for chiplets that compute different tasks with the same input image. In particular, future work encourages us to focus on multi-task learning encoders, as a hardened encoder can run a variety of different tasks without much area increase from different decoder heads, greatly increasing flexibility.

Other possible avenues for future exploration include hardening different model architectures, such as transformers, as well as increasing task flexibility by embedding Low-Rank Adaptation matrices into the hardened convolutional layers. For actual chip production, there are still open questions to resolve, including making the chip robust to glitches. By providing our results, we hope to foster community discussion and adoption.



\bibliographystyle{ACM-Reference-Format}

\begin{bibliography}{}
\end{bibliography}

\end{document}

%% file: graphs/weight-encoding.tex
\usetikzlibrary{matrix}
\begin{tikzpicture}
    \matrix[matrix of nodes, 
            nodes={draw, anchor=center, minimum size=1cm, text height=1.5ex, text depth=.25ex},
            column sep=0pt, row sep=0pt,
            xshift=-3cm] (m) {
        \node[fill=red!50]{}; & \node[fill=white]{}; & \node[fill=red!50]{}; & \node[fill=white]{}; & \node[fill=yellow!50]{}; & \node[fill=white]{}; & \node[fill=yellow!50] {}; & \node[fill=white] {}; \\
        \node[fill=orange!50] {}; & \node[fill=orange!50] {}; & \node[fill=white] {}; & \node[fill=white] {}; & \node[fill=blue!50] {}; & \node[fill=blue!50] {}; & \node[fill=white] {}; & \node[fill=white] {}; \\
        \node[fill=white] {}; & \node[fill=pink!50] {}; & \node[fill=pink!50] {}; & \node[fill=white] {}; & \node[fill=cyan!50] {}; & \node[fill=white] {}; & \node[fill=white] {}; & \node[fill=cyan!50] {}; \\
        \node[fill=white] {}; & \node[fill=white] {}; & \node[fill=purple!50] {}; & \node[fill=purple!50] {}; & \node[fill=green!50] {}; & \node[fill=green!50] {}; & \node[fill=white] {}; & \node[fill=white] {}; \\
    };

    \matrix[matrix of nodes, 
            nodes={draw, anchor=center, minimum size=1cm, text height=1.5ex, text depth=.25ex},
            column sep=0pt, row sep=0pt,
            yshift=0cm,xshift=4cm] (m_compressed) {
        \node[fill=red!50]{}; & \node[fill=red!50]{}; & \node[fill=yellow!50]{}; & \node[fill=yellow!50] {}; \\
        \node[fill=orange!50] {}; & \node[fill=orange!50] {}; & \node[fill=blue!50] {}; & \node[fill=blue!50] {}; \\ \node[fill=pink!50] {}; & \node[fill=pink!50] {}; & \node[fill=cyan!50] {}; & \node[fill=cyan!50] {}; \\
        \node[fill=purple!50] {}; & \node[fill=purple!50] {}; & \node[fill=green!50] {}; & \node[fill=green!50] {};\\
    };

    \matrix[matrix of nodes, 
            nodes={draw, anchor=center, minimum size=1cm, minimum width = 0.6cm, text height=1.5ex, text depth=.25ex,font=\Huge},
            column sep=0pt, row sep=0pt,
            yshift=0cm,xshift=8cm] (m_metadata) {
            0  & 2  & 0  & 2 \\
            0 & 1 & 0 & 1 \\
            1 & 2 & 0 & 3 \\
            2 & 3 & 0 & 1 \\
       };

\end{tikzpicture}

%% file: graphs/fig1.1.tex
\begin{circuitikz}
    \draw (-1,2) rectangle (1,3.5);
    \node at (0,2.75) {\Huge *};  

    \draw (-1,0) rectangle (1,1.5);
    \node at (0,0.75) {\Huge *};  

    \draw (-2,3.2) node[left] {\Large 0.520} -- (-1,3.2);
    \draw (-2,2.3) node[left] {\Large 0.048} -- (-1,2.3);

    \draw (-2,1.2) node[left] {\Large 0.280} -- (-1,1.2);
    \draw (-2,0.3) node[left] {\Large 0.035} -- (-1,0.3);

    \draw (1,2.75) -- (2.5,2.75) node[midway, above] {\Large 0.02496} -- (2.5,2.75);
    \draw (1,0.75) -- (2.5,0.75) node[midway, below] {\Large 0.0098} -- (2.5,0.75);

    \draw (2.5,2.75)  -- (2.5, 1.75);
    \draw (2.5,0.75) -- (2.5, 1.75);
    \draw (2.5,1.75) -- (3.5, 1.75);
    
    \draw (3.5,1) rectangle (5.5,2.5);
    \node at (4.5,1.75) {\Huge +};  

    \draw (5.5,1.75) -- (6.5,1.75) node[right] {\Large 0.03476};
\end{circuitikz}

%% file: graphs/fig1.2.tex
\begin{circuitikz}
    \draw (-1,2) rectangle (1,3.5);
    \node at (0,2.75) {\Huge *};  

    \draw (-1,0) rectangle (1,1.5);
    \node at (0,0.75) {\Huge *};  

    \draw (-2,3.2) node[left] {\Large 0.5} -- (-1,3.2);
    \draw (-2,2.3) node[left] {\Large 0.048} -- (-1,2.3);

    \draw (-2,1.2) node[left] {\Large 0.25} -- (-1,1.2);
    \draw (-2,0.3) node[left] {\Large 0.035} -- (-1,0.3);

    \draw (1,2.75) -- (2.5,2.75) node[midway, above] {\Large 0.0240} -- (2.5,2.75);
    \draw (1,0.75) -- (2.5,0.75) node[midway, below] {\Large 0.00875} -- (2.5,0.75);

    \draw (2.5,2.75)  -- (2.5, 1.75);
    \draw (2.5,0.75) -- (2.5, 1.75);
    \draw (2.5,1.75) -- (3.5, 1.75);
    
    \draw (3.5,1) rectangle (5.5,2.5);
    \node at (4.5,1.75) {\Huge +};  

    \draw (5.5,1.75) -- (6.5,1.75) node[right] {\Large 0.03275};
\end{circuitikz}

%% file: graphs/fig1.3.tex
\begin{circuitikz}
    \draw (-1,2) rectangle (1,3.5);
    \node at (0,2.75) {\Large $>>$};  

    \draw (-1,0) rectangle (1,1.5);
    \node at (0,0.75) {\Large $>>$};  

    \draw (-2,3.2) node[left] {\Large 1} -- (-1,3.2);
    \draw (-2,2.3) node[left] {\Large 0.048} -- (-1,2.3);

    \draw (-2,1.2) node[left] {\Large 2} -- (-1,1.2);
    \draw (-2,0.3) node[left] {\Large 0.035} -- (-1,0.3);

    \draw (1,2.75) -- (2.5,2.75) node[midway, above] {\Large 0.0240} -- (2.5,2.75);
    \draw (1,0.75) -- (2.5,0.75) node[midway, below] {\Large 0.00875} -- (2.5,0.75);

    \draw (2.5,2.75)  -- (2.5, 1.75);
    \draw (2.5,0.75) -- (2.5, 1.75);
    \draw (2.5,1.75) -- (3.5, 1.75);
    
    \draw (3.5,1) rectangle (5.5,2.5);
    \node at (4.5,1.75) {\Huge +};  

    \draw (5.5,1.75) -- (6.5,1.75) node[right] {\Large 0.03275};
\end{circuitikz}

%% file: graphs/fig1.4.tex
\begin{circuitikz}
        \draw (-2,2.3) node[left] {\Large 0.0481} -- (-1.5,2.3);

        \draw (-2,0.3) node[left] {\Large 0.035} -- (-1.5,0.3);

        \draw (-1.5, 2.75) -- (-1.5, 1.75);
        \draw (-1.5, -0.25) -- (-1.5, 0.75);

        \foreach \y in {2.75, 2.607, 2.464, 2.321, 2.178, 2.035, 1.892, 1.75} {
            \draw (-1.5,\y) -- (0,\y);
        }

        \foreach \y in {2.75, 2.607, 2.464, 2.321, 2.178, 2.035, 1.892} {
            \draw (0,\y) -- (1.5,\y-0.1429);
        }
        
        
        \draw (0,2.75+0.1429) -- (1.5,2.75);

        \draw (1.5, 2.75) -- (1.5, 1.75);

        
        \foreach \y in {2.75, 2.607, 2.464, 2.321, 2.178, 2.035, 1.892, 1.75} {
            \draw (-1.5,\y-2) -- (0,\y-2);
        }

        \foreach \y in {2.75, 2.607, 2.464, 2.321, 2.178, 2.035} {
            \draw (0,\y-2) -- (1.5,\y-2-2*0.1429);
        }

        \draw (0, 0.75+1*0.1429) -- (1.5, 0.75-0.1429);
        \draw (0, 0.75+2*0.1429) -- (1.5, 0.75);

        \draw (1.5, -0.25) -- (1.5, 0.75);

        \draw (1.5,2.3) -- (2.5,2.3) node[midway, above, xshift=0.1cm] {\Large 0.0240};
        \draw (1.5,0.3) -- (2.5,0.3) node[midway, below, xshift=0.2cm] {\Large 0.00875};
        \draw (2.5,2.3)  -- (2.5, 1.3);
        \draw (2.5,0.3) -- (2.5, 1.3);
        \draw (2.5,1.3) -- (3.5, 1.3);
    
        \draw (3.5,0.55) rectangle (5.5,2.05);
        \node at (4.5,1.3) {\Huge +};  

        \draw (5.5,1.3) -- (6.5,1.3) node[right] {\Large 0.03275};

    \end{circuitikz}

%% file: graphs/size-accuracy2.tex
\begin{tikzpicture}
\begin{axis}[
    title={Hardware Size vs Accuracy for CNN Models},
    title style={font=\large},
    xlabel={Size (mm$^2$)},
    ylabel={Top-1 Accuracy (\%)},
    xmin=20, xmax=15000,
    ymin=67, ymax=80,
    xmode=log,
    width=13cm,
    height=10cm,
    legend pos=north west
]

\addplot[only marks, mark=*, mark size=2pt] coordinates {
    (202.96, 71.8)        
    (145.08, 74.0)        
    (37.20, 67.7)         
    (2769.79, 76.0)       
    (1229.03, 69.8)       
    (10400.20, 73.3)      
    (5235.60, 77.0)       
    (1016.18, 69.8)       
    (97.51, 69.4)         
    (2725.69, 77.6)       
    (7722.89, 78.0)       
    (212.21, 73.5)        
};

\draw[red, dashed, thick] (axis cs: 850,67) -- (axis cs: 850,80) node[midway, above, rotate=90, yshift=2pt] {Reticle limit};

\node[anchor=south] at (axis cs:202.96, 71.8) {Mobilenet-V2};
\node[anchor=south] at (axis cs:145.08, 74.0) [xshift=5pt] {Mobilenet-V3-large};
\node[anchor=west] at (axis cs:37.20, 67.7) [xshift=5pt] {Mobilenet-V3-Small};
\node[anchor=north] at (axis cs:2769.79, 76.0) {Resnet-50};
\node[anchor=north] at (axis cs:1229.03, 69.8) [xshift=8pt] {Resnet-18};
\node[anchor=south] at (axis cs:10400.20, 73.3) [xshift=-5pt] {VGG16};
\node[anchor=north] at (axis cs:5235.60, 77.0) {Densenet-161};
\node[anchor=south] at (axis cs:1016.18, 69.8) [xshift=-8pt] {Googlenet};
\node[anchor=south] at (axis cs:97.51, 69.4) {Shufflenet};
\node[anchor=south] at (axis cs:2725.69, 77.6) [xshift=-5pt] {resnext50\_32x4d};
\node[anchor=south] at (axis cs:7722.89, 78.0) [xshift=-5pt,yshift=2pt] {wide\_resnet50\_2};
\node[anchor=south] at (axis cs:212.21, 73.5) [xshift=5pt] {MnasNet};

\end{axis}
\end{tikzpicture}

%% file: graphs/sparsity-vs-accuracy.tex
\begin{tikzpicture}
\begin{axis}[
    width=12cm,
    height=7cm,
    scale only axis,
    xlabel={\Huge Sparsity (\%)},
    ylabel={\Huge Accuracy (\%)},
    xmin=0, xmax=80,
    ymin=0, ymax=90,
    xtick={0, 20, 40, 60, 80},
    ytick={20, 40, 60, 80},
    grid=both,
    legend pos=south west
]

\addplot[
    color=blue,
    mark=*,
    thick
]
coordinates {
    (0, 68.6) 
    (20, 68.462)
    (40, 67.986)
    (60, 66.056)
    (63, 64.5)
    (66, 61.4)
    (80, 56.676)
};
\addlegendentry{Top-1 Accuracy}

\addplot[
    color=red,
    mark=square*,
    thick
]
coordinates {
    (0, 88.6) 
    (20, 88.381996)
    (40, 87.897995)
    (60, 86.992)
    (63, 85.97)
    (66, 84.1)
    (80, 81.392)
};
\addlegendentry{Top-5 Accuracy}

\end{axis}
\end{tikzpicture}

%% file: graphs/area_vs_sparsity.tex
\begin{tikzpicture}
\begin{axis}[
    xlabel={\Huge Sparsity (\%)},
    ylabel={\Huge Area ($\mu m^2$) },
    grid=both,
    xtick={0,20,40,60,80},
    xmin=0, xmax=80,
    ymajorgrids=true,
    width=12cm,   
    height=7cm,    
    scale only axis
]

\addplot coordinates {(0, 548.977184)
                    (20, 439.1817472)
                    (40, 329.3863104)
                    (60, 219.5908736)
                    (80, 109.7954368)};

\end{axis}
\end{tikzpicture}

%% file: graphs/throughput_vs_sparsity.tex
\begin{tikzpicture}
\begin{axis}[
    width=13cm,
    height=6cm,
    scale only axis,
    xlabel={\Huge Sparsity (\%)},
    ylabel={\Huge Throughput ($10^6$ images/sec)},
    domain=0:100,
    samples=201,
    grid=both,
    xtick={0,10,20,30,40,50,60,70,80,90},
    xmin=0,xmax=80,
    ytick={0,0.1,0.2,0.3,0.4,0.5,0.6,0.7,0.8,0.9,1.0,1.1,1.2,1.3,1.4},
    yticklabels={,, , , ,0.5, , , , ,1.0, , , ,},
    ymajorgrids=true,
    width=12cm,
    height=7cm
]

\addplot[blue] {
    (1e3 / max(
        ((224^2 * 3 + 1000 * 4) * floor(850 / (550 * abs(1 - 0.01 * x)))) /
        (900 * (550 * abs(1 - 0.01 * x)) / 800),
        3300)
    ) * floor(850 / (550 * (1 - 0.01 * x)))
};

\addplot[
    blue,
    only marks,
    mark=*,
    samples at={0,10,20,30,40,50,60,70,80}
] {(1e3 / max(
        ((224^2 * 3 + 1000 * 4) * floor(850 / (550 * abs(1 - 0.01 * x)))) /
        (900 * (550 * abs(1 - 0.01 * x)) / 800),
        3300)
    ) * floor(850 / (550 * (1 - 0.01 * x)))};


\end{axis}
\end{tikzpicture}

%% file: graphs/transfer-learning.tex
\begin{tikzpicture}
\begin{axis}[
    ybar,
    bar width=0.2cm,
    width=12cm,
    height=6cm,
    ylabel={\Large Accuracy (\%)},
    symbolic x coords={Flowers102, OxfordPets, Food101, CIFAR10, CIFAR100},
    xtick=data,
    ymin=0, ymax=100,
    enlarge x limits=0.2,
    legend style={at={(0.5,-0.15)}, anchor=north, legend columns=3},
    every node near coord/.append style={font=\Large},
    grid=major,
]

\addplot+[style={fill=blue!60}] coordinates {(Flowers102, 81.22) (OxfordPets, 91.03) (Food101, 60.4) (CIFAR10, 76.1)(CIFAR100, 78.7)};
\addplot+[style={fill=orange!80}] coordinates {(Flowers102, 79.59) (OxfordPets, 88.59) (Food101, 56.9) (CIFAR10, 71.3)(CIFAR100, 58.9)};
\addplot+[style={fill=green!70}] coordinates {(Flowers102, 77.83) (OxfordPets, 84.78) (Food101, 56.9) (CIFAR10, 79.1) (CIFAR100, 59.2)};

\legend{Original MobileNetV2, Q3.5 HaShiFlex, Q3.5 HaShiFlex $60\%$ Pruned}
\end{axis}
\end{tikzpicture}